\documentclass[showpacs,preprint,preprintnumbers,amsmath,amssymb]{revtex4}
\begin{document}
\title{ Consequences of self-consistency violations in Hartree-Fock
random-phase approximation
calculations of the nuclear breathing mode energy}
\author{B. K. Agrawal and S. Shlomo}
\address{Cyclotron Institute, Texas A\&M University,
College Station, TX 77843, USA}

\begin{abstract}
We provide for the first time accurate assessments  of the consequences
of violations of self-consistency in the Hartree-Fock based random phase
approximation (RPA) as commonly used  to calculate the energy $E_c$
of the  nuclear breathing mode.  Using several Skyrme interactions we
find  that the  self-consistency violated by ignoring the spin-orbit
interaction in the RPA calculation causes a spurious enhancement of
the breathing mode energy for spin unsaturated systems.  Contrarily,
neglecting the Coulomb interaction in the RPA or performing the RPA
calculations in the TJ scheme underestimates the  breathing mode energy.
Surprisingly, our results for the $^{90}$Zr and $^{208}$Pb nuclei for
several Skyrme type effective nucleon-nucleon interactions having a
wide range of nuclear matter incompressibility ($K_{nm} \sim 215 -
275$ MeV) and symmetry energy ($J \sim 27 - 37$ MeV)  indicate that
the net uncertainty ($\delta E_c \sim  0.3$ MeV) is comparable to the
experimental one.

\end{abstract}

\pacs{21.60.Jz, 24.30.Cz, 21.65.+f}
\maketitle
The Hartree-Fock (HF) based random phase approximation (RPA) provides
a microscopic description of the nuclear compressional modes. The
most special of these compressional modes is the isoscalar giant
monopole resonance (ISGMR) also referred to as the breathing mode. The
centroid energy $E_c$ of the ISGMR enables one to determine the value of
nuclear matter incompressibility coefficient $K_{nm}$ which plays an
important role in understanding a wide variety of phenomena ranging
from heavy-ion collision to supernova explosions. Recent experimental
data \cite{Youngblood99} for the $E_c$ in heavy nuclei have rather small
uncertainties ($\sim  0.1 - 0.3$ MeV).  Since, the uncertainty $\delta
E_c$ associated with $E_c$ is approximately related to the uncertainty  $\delta K_{nm}$ in $K_{nm}$ by,
\begin{equation}
\label{delk}
\frac{\delta K_{nm}}{K_{nm}}=2\frac{\delta E_c}{E_c},
\end{equation}
the value of  $\delta K_{nm}$ is only about 10 MeV, for $K_{nm} = 250$ MeV
and $E_c = 14.17\pm 0.28$ MeV for the $^{208}$Pb nucleus.  On the
theoretical side, most of the results obtained within the  HF-RPA theory,
as employed for the determination of $E_c$, are plagued by the lack of
self-consistency, see however, Ref. \cite{Reinhard92}.  Self-consistency
is violated due to the neglect of the spin-orbit and Coulomb terms in the
particle-hole interaction used in the RPA calculations.  Furthermore,
some of the RPA calculations are performed in the TJ (isospin) scheme.
Uncertainties in the value of $E_c$ calculated in the HF-RPA approach
arise also  due to various numerical approximations such as  the use
of  very small particle-hole excitation energies  and the  introduction
of the smearing parameter $\Gamma$ to smoothen the strength function.
In Ref. \cite{Agrawal03}, we have investigated in detail the effects
of these numerical approximations and emphasized the need to perform a
highly accurate calculation of the strength function. So, it is of utmost
importance that in  order to determine an accurate value of $K_{nm}$
one must have an accurate knowledge of the effects of violations of
self-consistency on the centroid energy of the ISGMR.  We may also point
out that modifying the particle-hole interaction in an ad hoc manner
in such a way that the spurious state associated with  the center of
mass motion appears at zero energy \cite{Shlomo75} does not restore the
self-consistency (see also Ref. \cite{Agrawal03}).

In this work we provide for the first time accurate assessments of the
effects of violations of self-consistency on the constrained and the scaling
energies,
\begin{equation}
\label{e0-13}
E_{con}=\sqrt{\frac{m_1}{m_{-1}}},
\qquad {\rm and}\qquad
E_s=\sqrt{\frac{m_3}{m_1}},
\end{equation}
of the ISGMR. Here  
\begin{equation}
\label{mk}
m_k=\int_{0}^{\infty}\omega^k  S(\omega) d\omega 
\end{equation}
is the energy moments of the strength function
\begin{equation}
S(\omega)=\sum_n\left |\langle n\mid f\mid 0\rangle\right |^2
\delta(\omega-\omega_n),
\end{equation}
for the monopole operator
$ f(r)=\sum_{i=1}^A r^2_i.$

In the following the fully self-consistent and highly accurate values of
the constraint energy, $E_{con}$, and the scaling  energy, $E_s$, are
obtained by calculating the energy moments $m_k$ for $k= -1, 1 $ and 3
using the constrained HF (CHF), the double commutator and the generalized
scaling approaches, respectively \cite{Bohigas79}. The self-consistent
values of $E_{con}$ and $E_s$ are then compared with those obtained
within the HF-RPA approach \cite{Agrawal03,Shlomo75} in order to assess
the effects of each type of violation of self-consistency.  It may be
pointed out that very recently  in Ref. \cite{Colo} some preliminary
attempts are made to estimate  the effects of self-consistency violation
on the value of $E_{con}$.  However, $E_s$ was not considered and the
different components contributing to the self-consistency violations
are not dealt separately. We show here that the appropriate knowledge
of the effects of  each of these components on the ISGMR energies  are
instrumental in validating the HF-RPA  calculations.

We consider here the HF solution for the Hamiltonian,
\begin{equation}
H = \sum_{i=1}^{A}  T_i +
\frac{1}{2}\sum_{i,j=1}^{A}V_{ij}+\frac{1}{2}\sum_{i,j=1}^Z V_{ij}^C,
\end{equation}
where $T$ is the kinetic energy operator, $V^C$ is the
Coulomb interaction   and  $V$ is the effective two-body
interaction  of the Skyrme type  \cite{Vautherin72},
\begin{eqnarray}
\label{v12}
&V_{12}&= t_0\left (1+x_0 P_{12}^\sigma\right )\delta({\bf r}_1-{\bf
r}_2)\nonumber\\
&&+\frac{1}{2}t_1\left (1+x_1 P_{12}^\sigma\right ) 
\times \left[\overleftarrow{k}_{12}^2\delta({\bf r}_1-{\bf r}_2)+\delta({\bf
r}_1-{\bf r}_2)\overrightarrow{k}_{12}^2\right] \nonumber\\
&&+t_2\left (1+x_2 P_{12}^\sigma\right )\overleftarrow{k}_{12}\delta({\bf r}_1-{\bf r}_2)\overrightarrow{k}_{12} \nonumber\\
&&+ \frac{1}{6}t_3\left (1+x_3 P_{12}^\sigma\right )\rho^\alpha\left(\frac{{\bf r}_1+{\bf r}_2}{2}\right )
\delta({\bf r}_1-{\bf r}_2) \nonumber\\
&&+iW_0\overleftarrow{k}_{12} \delta({\bf r}_1-{\bf r}_2)(\overrightarrow{\sigma_1}+
\overrightarrow{\sigma_2})\times \overrightarrow{k}_{12},
\end{eqnarray}
where $P_{12}^\sigma$ is the spin exchange operator,$\overrightarrow{\sigma}_i$ is the Pauli spin operator,
$\overrightarrow{k}_{12} = -i( \overrightarrow{\nabla}_1-\overrightarrow{\nabla}_2)/2$
 and 
$\overleftarrow{k}_{12} = -i(\overleftarrow{\nabla}_1-\overleftarrow{\nabla}_2)/2\, .$
Here, the   right and left arrows indicate that the momentum operators act
on the right and  on the left,  respectively.  The parameters of the
Skyrme force $t_i$, $x_i$, $\alpha$ and $W_0$  are obtained by fitting
the HF results to the experimental data for the bulk properties of
finite nuclei.  

Once the HF equations are solved, the $m_1$
and $m_3$ can be expressed in terms of the ground state density  $\rho$,
kinetic energy density $\tau$ and spin density ${\bf J}$. The value of
$m_1$ can be obtained from the corresponding double commutator \cite{Bohr75} 
and is given as,
\begin{equation}
\label{m1} m_1=2\frac{\hbar^2}{m}\langle r^2\rangle,
\end{equation}
where, 
\begin{equation} \langle r^2\rangle = \int r^2\rho(r) d{\bf r}.
\end{equation} 
The cubic moment $m_3$ can  be evaluated in the generalized scaling
approach and it is given by\cite{Bohigas79,Lipparini89},
\begin{eqnarray}
\label{m3} 
m_3&=&\frac{1}{2}\left (\frac{2\hbar^2}{m}\right )^2 \left [
2T+6E_{\delta}+20(E_{fin}+E_{so}) +\right .\nonumber\\
&&\left . (3\alpha + 2)(3\alpha+3)E_\rho\right ], 
\end{eqnarray}
where, 
\begin{eqnarray} T&=&\int \frac{\hbar}{2m}  \tau
d{\bf r},\\ \label{e-del}
E_\delta &=& \int \left [\frac{3}{8}t_0\rho^2
- \frac{1}{4}t_0(x_0+\frac{1}{2})\rho_1^2\right ] d{\bf r}\\
\label{e-fin} E_{fin}&=&\int\left [  a_0\rho\tau + a_1\rho_1\tau_1+
c(\nabla\rho)^2+d(\nabla\rho_1)^2\right ]d{\bf r}\\ \label{e-rho}
E_\rho&=& \int\left[ \frac{3}{48}t_3\rho^{\alpha+2}-\frac{1}{24}t_3(x_3+
\frac{1}{2})\rho^\alpha\rho_1^2\right ]d{\bf r}\\ \label{e-so}
E_{so}&=& -\frac{1}{2}W_0\int \left [\rho{\bf \nabla\cdot
J}+\rho_p{\bf \nabla\cdot J}_p+\rho_n{\bf\nabla \cdot J}_n\right
]d{\bf r} 
\end{eqnarray} 
with $\rho_1=\rho_n-\rho_p$, $\tau_1
= \tau_n-\tau_p$,  $a_0=\frac{1}{16}(3t_1+(5+4x_2)t_2)$,
$a_1=\frac{1}{8}t_2(x_2+\frac{1}{2})-t_1(x_1+\frac{1}{2})$,
$c=\frac{1}{64}(9t_1-(5+4x_2)+t_2)$ and
$d=-\frac{1}{32}(3t_1(x_1+\frac{1}{2})+t_2(x_2+\frac{1}{2}))$.
As described in
detail in Ref. \cite{Bohigas79}, $m_{-1}$ can be evaluated via 
the CHF approach and is given as, 
\begin{equation} \label{m-1} m_{-1}=\left
. \frac{1}{2}\frac{d}{d\lambda}\langle r^2_\lambda \rangle \right
|_{\lambda=0}=\left . \frac{1}{2}\frac{d^2}{d\lambda^2}\langle
H_\lambda\rangle \right |_{\lambda=0} 
\end{equation} 
where, 
$ \langle O_\lambda\rangle = \langle \Phi_\lambda\left
| O\right |\Phi_\lambda \rangle, 
$ 
with $\Phi_\lambda$ being the HF solution for the
constrained  Hamiltonian,
$ H_\lambda = H - \lambda f.  $

The corresponding HF-RPA values of $m_k$ 
are obtained from Eq. (\ref{mk}) using,
\begin{equation} 
\label{se-rpa}
S(\omega)=\frac{1}{\pi} Im \left [Tr
(fG(\omega)f)\right ] 
\end{equation} 
with $G(\omega)$ being the RPA Green's function \cite{Shlomo75,Agrawal03}.
The differences between the values of $m_k$ and consequently the difference
\begin{equation}
\delta E  = E(RPA) - E(SC),
\end{equation}
between the energies obtained within the HF-RPA and  the self-consistent (SC) values obtained 
within the HF approach  are due to the
violation of  self-consistency  in the HF-RPA. 
Since we are unable to obtain a fully self-consistent value for $m_0$,
we only provide the value of the centroid energy  
\begin{equation}E_c = \frac{m_1}{m_0},
\end{equation}
obtained within the HF-RPA. Thus we use  $\delta
E_{con}$  and $\delta E_s$ as measures of $\delta E_c$.                                                     
We point out  that we find $E_c \approx E_{con}$.

Before we present the main results, some remarks about the numerical
accuracies are in order.  We carry out the HF calculations using a box
of 18 fm and a mesh size of $\delta r = 0.02$ fm.  To determine $m_{-1}$
given by Eq. (\ref{m-1}) with an accuracy of 0.1\% we calculate the
first derivative of $\langle r^2_\lambda\rangle $ using a five-point
formula  with the increment $\delta \lambda = 0.02$. In order to limit
the spurious enhancement in $m_k$ ( particularly $m_3$), due to the
Lorentzian smearing of the RPA Green's function, to below 0.2\%, we carry
out the continuum RPA (CRPA) and the discretized RPA (DRPA) calculations
using the very small smearing parameter of $\Gamma/2 = 5\times 10^{-2}$
MeV and the maximum energy of $4E_c$ in the integral of Eq. (\ref{mk}).
In Ref. \cite{Agrawal03} we show that in order to calculate the centroid
energy accurate to within 0.1 MeV, one must set the cut-off for the
p-h excitations  $E_{ph}^{max} > 400$ MeV.  In what follows, we shall
present our DRPA results obtained with $E_{ph}^{max} = 950$ MeV.

In Table \ref{fsc} we demonstrate the accuracy of our calculations by
employing a simplified Skyrme interaction \cite{Agrawal03} with $t_0=
-1800$ MeVfm$^3$, $t_3= 12871 $ MeVfm$^4$  and $\alpha=1/3$. We consider
the $^{80}$Zr nucleus and  keep
the spin-orbit and the Coulomb interactions switched off in the HF
as well as in RPA calculations, so that the HF-RPA calculations are
fully self-consistent. Self-consistent (SC)   values of $m_k$ and the
energies  $E_{con}$ and $E_s$ shown in the first row of this table
are obtained using Eqs. (\ref{e0-13}), (\ref{m1}), (\ref{m3}) and
(\ref{m-1}). These values are accurate to within $0.1\%$.  We compare
these highly accurate results with the corresponding ones obtained
within the HF-RPA approach. The CRPA and DRPA results presented in
Table \ref{fsc} are calculated by integrating the RPA strength function
appearing in Eq. (\ref{mk}) up to the different values of the  maximum
energy $\omega_{max}$. One can easily verify that the CRPA as well
as DRPA results are quite close to the highly accurate  SC results
(first row).  For instance, the maximum uncertainty is associated with
the  $E_s$ and it is about $0.3\%$ for $\omega_{max} = 100$ MeV.  Since,
the RPA  energies $E_{con}$, $E_c$ and $E_s$ obtained by integrating
Eq. (\ref{mk}) up to  $\omega_{max} = 100$ and 120 MeV are very close,
we shall adopt in the following the value of $\omega_{max} \approx 4E_c$
for the RPA calculations.

Having demonstrated in Table \ref{fsc} the  high accuracy of our
numerical calculations, we now consider the effects of the violations
of self-consistency.  To investigate effects of each of the components
contributing to the self-consistency violations, we carry out the
calculations for some specific nuclei using several interactions
\cite{Giai81,Bartel82,Chabanat98,Au03}.  We denote  by $\theta_{ls}
(\theta_c) = 0$ or 1 that the spin-orbit (Coulomb) interaction is
turned off or on in the HF calculations. In the RPA calculations the
p-h spin-orbit and Coulomb interactions are always absent.  Thus, the
HF-RPA calculations for symmetric nuclei (N = Z)  with $\theta_{ls} =
\theta_c = 0$ are fully self-consistent.  The effects of ignoring the
p-h spin-orbit interaction on the ISGMR energies can be singled out by
carrying out the calculations for symmetric nuclei with $\theta_{ls} =
1$ and $ \theta_c = 0$. Similarly the effects due to the neglect of p-h
Coulomb interaction can be estimated by performing the calculations
for the cases with  $\theta_{ls}=0$ and $\theta_c = 1$. The HF-RPA
calculations for asymmetric nuclei with $\theta_{ls} = \theta_c = 0$
yields the effect of working in the TJ scheme.  In Table \ref{s-o} we
present the results  obtained using the SGII interaction \cite{Giai81}
for  several nuclei for different combinations of $\theta_{ls}$ and
$\theta_{c}$. Results for other Skyrme interactions exhibit similar
features.  The $^{40,60}$Ca and $^{80,110}$Zr nuclei considered here
are spin saturated systems. But, $^{56}$Ni and $^{100}$Sn are spin
unsaturated nuclei  having $1f_{7/2}$ and $1g_{9/2}$ orbits as the
last occupied ones, respectively.  As seen in Table \ref{fsc}, it is
once again evident that the fully self-consistent HF-RPA calculations
for the $^{40}$Ca and $^{80}$Zr nuclei ($\theta_{ls} = \theta_c = 0$
case) yield  small values for $\delta E_{con}$ and $\delta E_s$ ($ <
0.1$ MeV).  Considering the results for the cases with $\theta_{ls} =
1$, we find that  for the spin saturated nuclei the ISGMR energy do
not get altered much even if the p-h  spin-orbit interaction is not
included. However, in case of the spin unsaturated nuclei, the effect
of the neglect of the p-h  spin-orbit interaction in the RPA leads to
a significant spurious enhancement of the ISGMR energy.  Following Eq.
(\ref{delk}), one can easily verify that such enhancements in the ISGMR
energies are equivalent to overestimating the value of  $K_{nm}$ by about
30-40 MeV, which is unreasonably large in view of the current accuracy
of the experimental data.  Let us now focus on the results obtained for
$^{60}$Ca and $^{110}$Zr nuclei with  $\theta_{ls} = \theta_c = 0$ and
$^{40}$Ca and $^{80}$Zr nuclei with  $\theta_{ls} = 0$ but $\theta_c =
1$. Compilation of these results enables us to investigate the effects
of working in the TJ scheme as well as the neglect of the Coulomb term
in the  p-h interaction used in the RPA calculations. One can conclude
from Table \ref{s-o} that both of these factors, contributing to the
violation of self-consistency,  tend to underestimate the ISGMR energy
compared with its SC value.

So far we have seen in a systematic manner how the inconsistency
in implementing the HF-RPA approach introduces uncertainty in  the
ISGMR energy.  Considering the results of   Tables \ref{s-o}, one may
expect that the total effects of the self-consistency violation on the
ISGMR energy may be smaller. Because, the absence of p-h spin-orbit
interaction tends to increase the ISGMR energy while working in TJ
scheme or ignoring the Coulomb term in p-h interaction used  in the
RPA tends to underestimate the ISGMR energy.  As an illustration, we
present in Table \ref{zr-pb} the results for $^{90}Zr$ and $^{208}$Pb
nuclei which are  widely used to extract the value of $K_{nm}$ from the
ISGMR energies. These calculations are carried out for several Skyrme
interactions such as SGII, SkM$^*$, SLy4, SK255 and SK272 \cite{Giai81,
Bartel82, Chabanat98, Au03}.  The choice of these interactions cover a
wide range of incompressibility coefficient $K_{nm} = 214 - 272$ MeV,
symmetry energy coefficient $J = 27 - 37$ MeV and the spin-orbit strength
$W_0 = 95 - 130$ MeVfm$^5$.  We see that in the case of $^{90}$Zr, the
ISGMR energy $E_{con}$ obtained in DRPA agrees with the corresponding
self-consistent values to  within 0.3 MeV.  For the $^{208}$Pb nucleus,
we have $\mid \delta E_{con} \mid \leq 0.4$ MeV.  For both nuclei,
the behaviour of $\delta  E_s$ is similar to that of $\delta E_{con}$
except for the  fact that former one is larger by about 0.2 MeV.  Thus,
the net uncertainties in the ISGMR energies due to the self-consistency
violations are reasonably small for the $^{90}$Zr and $^{208}$Pb nuclei
and are  comparable to that attained in recent experiments.  Of course,
this may not be the case for  spin saturated nuclei near the  drip line;
the inconsistency in the HF-RPA calculations introduced by the Coulomb
interaction and the asymmetry in these nuclei may overshadow the effect
of neglecting the p-h spin-orbit interaction.

In summary, we have  investigated in detail the effects of the
violations of the self-consistency  in the HF-RPA calculation of  
the ISGMR energies $E_{con}$ and $E_s$. In particular, we considered
the self-consistency violations caused  by ignoring the spin-orbit
and Coulomb terms in the p-h interaction and by carrying out the RPA
calculations in the TJ scheme. We performed  the HF-RPA calculations for
the ISGMR energies for several nuclei with the SGII Skyrme interaction
and demonstrated that ignoring the spin-orbit term in the p-h interaction
gives rise to a spurious enhancement in the values of $E_{con}$ and $E_s$
for spin unsaturated nuclei. On the contrary, neglect of the Coulomb
term in the p-h interaction and performing the RPA calculations in the
TJ scheme underestimates the ISGMR energies.  Finally,  we calculated
the ISGMR energies for the $^{90}$Zr and $^{208}$Pb nuclei for the five
different Skyrme interactions SGII, SkM$^*$, SLy4, SK255 and SK272 and
show that in these nuclei, widely used to extract the value of $K_{nm}$,
the various elements contributing to the self-consistency violations tend
to counterbalance their effects leading to an uncertainty of about $0.1
- 0.4$ MeV in the values of ISGMR energies, which is quite acceptable in
view of the accuracy of the experimental data currently available.

This work was supported in part by the US Department of Energy under grant
\# DOE-FG03-93ER40773 and the National Science Foundation under grant
$\#$PHY-0355200.

\begin{table}[th]
\caption{\label{fsc}
Comparison of self-consistent (SC) values of the moments $m_k$ (in
MeV$^k$fm$^4$) and the ISGMR energies (in MeV)  for the $^{80}$Zr
nucleus calculated in the HF approach with the ones obtained within the
continuum and discretized RPA using the smearing parameter $\Gamma =
0.01$ MeV.  The RPA values are obtained by integrating Eq. (\ref{mk})
over the energy range $0-\omega_{max}$ (in MeV). The calculations
are performed using a simplified Skyrme interaction with $t_0= -1800$
MeVfm$^3$, $t_3= 12871 $ MeVfm$^4$  and $\alpha=1/3$. }
\begin{ruledtabular}
\begin{tabular}{|ccccccccc|}
\hline
& $\omega_{max}$&$m_{-1}$&$ m_0$&$m_1$&$m_3$& $E_{con}$&  $E_c$& $E_s$\\
\hline
SC&   &  185.1&   & 96346& 54490227&   22.81 &   & 23.78\\
CRPA& 80&   185.0&4189& 96229& 53549803&22.80&22.97&23.59\\
CRPA& 100&   185.1&4190& 96289& 54024448&22.81&22.98&23.69\\
CRPA& 120& 185.1&4190& 96303& 54182661&22.81&22.98&23.72\\
DRPA& 80&   185.4&4194& 96359& 53721598&22.80&22.97&23.61\\
DRPA& 100&   185.4&4196& 96426& 54292408&22.80&22.98&23.73\\
DRPA& 120&185.4&4196& 96449& 54544736&22.81&22.99&23.78\\
\hline
\end{tabular}
\end{ruledtabular}
\end{table}
\vspace*{3 in}
\begin{table}
\caption{\label{s-o}
Comparison between the  SC and DRPA results for the ISGMR energies (in
MeV)  for several  nuclei obtained for the SGII interaction \cite{Giai81}.
The notation $\theta_{ls} (\theta_c) = 1$ or 0 indicates that the HF
calculations are performed with or without the inclusion of the spin-orbit
(Coulomb)  interaction.  Note that the p-h interaction used in the RPA
calculations does not include the spin-orbit and Coulomb terms.}
\begin{ruledtabular}
\begin{tabular}{|ccccccccrr|}
\hline
\multicolumn{3}{|c}{}&
\multicolumn{3}{c}{DRPA}&
\multicolumn{2}{|c}{SC}&
\multicolumn{2}{c|}{}\\
\cline{4-8}
Nucleus&$\theta_{ls}$ &$\theta_{c}$
&$E_{con}$&  $E_c$& $E_s$&$E_{con}$& $E_s$
&$\delta  E_{con}$ & $\delta E_s$ \\
\hline
$^{40}$Ca&0 &0&22.53&22.76&23.85 &22.60&   23.92& -0.07& -0.07\\
&1   &0&22.31&22.63&23.96 & 22.58&   23.93& -0.27& 0.03\\
&  0 &1  &20.97&21.24&22.44  & 21.38  & 22.69& -0.41& -0.25\\
\hline
 $^{80}$Zr& 0  &0&19.57&19.69&20.26 & 19.60&  20.32&-0.03&-0.06\\
& 1  &0&19.49&19.65&20.36& 19.58 &20.33&-0.09&0.03\\
& 0 &1& 17.38&17.54&18.20 &    17.81  & 18.63& -0.43& -0.43\\
\hline
$^{56}$Ni& 1   &0& 23.14&23.35&24.44 & 21.49&  22.41& 1.65& 2.03\\
$^{100}$Sn  & 1  & 0& 20.16& 20.27& 20.95 &19.04 & 19.57&1.12& 1.38\\
$^{60}$Ca&0 & 0 &16.08&16.81&19.13 & 17.32&   19.57& -1.24& -0.44\\
$^{110}$Zr&0& 0&15.64&16.04&17.35&  16.33&   17.52& -0.69& -0.17\\
\end{tabular}
\end{ruledtabular}
\end{table}
\begin{table}
\caption{\label{zr-pb} 
Comparison between the SC and the DRPA results for the ISGMR energies
(in MeV) for the $^{90}$Zr and $^{208}$Pb nuclei.  The DRPA results are
obtained using the smearing parameter $\Gamma = 0.01$ MeV and the values
of $\omega_{max} = 70$ and  50 MeV  for the $^{90}$Zr and $^{208}$Pb
nuclei, respectively.  Note that the values of $\delta E_{con}$ are
comparable to the experimental uncertainties. }
\begin{ruledtabular}
\begin{tabular}{|cccccccrr|}
\hline
\multicolumn{2}{|c}{}&
\multicolumn{3}{c}{DRPA}&
\multicolumn{2}{c}{SC}&
\multicolumn{2}{c|}{}\\
\cline{3-7}
Nucleus& Int. &$E_{con}$& $E_c$& $E_s$ &$E_{con}$ &
$E_s$&
 $\delta E_{con}$ & $\delta E_s$\\ 
\hline
$^{90}$Zr&SGII&17.93&18.05&18.70 & 17.70& 18.30 & 0.23&0.40\\

 & SkM$^*$&17.72&17.85&18.55 & 17.43& 18.00& 0.29& 0.55\\
& SLy4& 18.19&18.33&19.07& 17.86& 18.46& 0.33& 0.61\\
& SK255&18.83&18.97&19.64  &18.87& 19.57 &- 0.04&0.07\\

& SK272&19.55&19.70&20.45& 19.49&   20.28 & 0.06& 0.17\\
\hline
 $^{208}$Pb&SGII &13.53&13.64&14.22 &13.42& 13.94 & 0.11& 0.28\\

  &  SkM$^*$&13.46&13.61&14.31 &13.36  & 13.88 & 0.10& 0.43\\

  &  SLy4& 14.04&14.16&14.80& 13.68&   14.22 &  0.36& 0.58\\

& SK255&13.83&13.98&14.65 & 14.26&   14.88&- 0.43&-0.23\\

& SK272&14.42&14.58&15.28 &14.73&  15.42 &- 0.31&-0.14\\
\hline
\end{tabular}
\end{ruledtabular}
\end{table}

\begin{thebibliography}{99}
\bibitem{Youngblood99} D. H. Youngblood, H. L. Clark and Y. W. Lui, Phys.
Rev. Lett. {\bf 82}, 691 (1999).
\bibitem{Reinhard92} P. G. Reinhard, Ann. Phys. (Leipzig) 1, 632 (1992)
and private communication.
\bibitem{Agrawal03} B. K. Agrawal, S. Shlomo and A. I. Sanzhur, Phys. Rev.
C {\bf 67},  034114 (2003).
\bibitem{Shlomo75} S.~Shlomo and G.~F.~Bertsch, Nucl. Phys. {\bf A243}, 
507 (1975).
\bibitem{Bohigas79} O. Bohigas, A. M. Lane and J. Martorell, Phys. Rep.
{\bf 51}, 267 (1979).
\bibitem{Colo}G. Colo and Nguyen Van Giai, Nucl. Phys. {\bf A731}, 15c
(2004).
\bibitem{Vautherin72} D. Vautherin and  D. M. Brink, Phys. Rev. C {\bf  5}, 626,(1972).
\bibitem{Bohr75} A.~Bohr and B.~Mottelson, {\em Nuclear Structure} 
(Benjamin, London, 1975), Vol. II, Chap. 6. 
\bibitem{Lipparini89} E. Lipparini and S. Stringari, Phys. Rep {\bf 175},
103 (1989).
\bibitem{Giai81} Nguyen Van Giai and H. Sagawa, Phys. Lett. B {\bf 106},
379 (1981).
\bibitem{Bartel82} J. Bartel, P. Quentin, M. Brack, C.
Guet and H. -B. Hakansson, Nucl. Phys. {\bf A386}, 79
(1982).
\bibitem{Chabanat98} E. Chabanat, P. Bonche, P. Haensel, J. Meyer and
R. Schaeffer,  Nucl. Phys. A635, 231 (1998).

\bibitem{Au03} B. Agrawal, S. Shlomo and V. Kim Au, Phys. Rev. C{\bf 68},
031304(R)  (2003).
\end{thebibliography}
\end{document}